%% file: cvpr21.tex
\documentclass[final]{cvpr}

\usepackage{times}
\usepackage{epsfig}
\usepackage{graphicx}
\usepackage{amsmath}
\usepackage{amssymb}
\usepackage{subfigure}
\usepackage{amsmath}
\usepackage{amssymb}
\usepackage{multirow}
\usepackage{algorithm}
\usepackage{algorithmic}
\usepackage{color}
\usepackage{makecell}
\usepackage{booktabs}

\usepackage[pagebackref=true,breaklinks=true,colorlinks,bookmarks=false]{hyperref}

\def\cvprPaperID{3078} 
\def\confYear{CVPR 2021}

\newcommand{\covid}{COVID-19\xspace}
\newcommand{\method}{TAN\xspace}
\newcommand{\methodfull}{Task Adaptation Network\xspace}

\newcommand{\flatten}{\textit{Flatten}\xspace}
\newcommand{\gram}{\textit{Gram}\xspace}
\newcommand{\source}{$\mathcal{S}$\xspace}
\newcommand{\target}{$\mathcal{T}$\xspace}
\newcommand{\pivot}{$\mathcal{P}$\xspace}
\newcommand{\vecx}{\mathbf{x}\xspace}
\newcommand{\vecy}{\mathbf{y}\xspace}

\begin{document}

\title{Learning Invariant Representations across Domains and Tasks}

\author{Jindong Wang$^{1\ast}$, Wenjie Feng$^2$, Chang Liu$^1$, Chaohui Yu$^3$, Mingxuan Du$^4$, \\Renjun Xu$^5$, Tao Qin$^1$, Tie-Yan Liu$^1$\\
	$^1$ Microsoft Research, Beijing, China   $^2$ Institute of Data Science, NUS, Singapore\\
	$^3$ Alibaba DAMO Academy $^4$ Hefei University of Technology $^5$ Zhejiang University\\
	{\tt\small \{jindong.wang, changliu\}@microsoft.com, wenjie.feng@nus.edu.sg, huakun.ych@alibaba-inc.com}
}

\maketitle

\begin{abstract}
Being expensive and time-consuming to collect massive COVID-19 image samples to train deep classification models, transfer learning is a promising approach by transferring knowledge from the abundant typical pneumonia datasets for COVID-19 image classification. However, negative transfer may deteriorate the performance due to the feature distribution divergence between two datasets and task semantic difference in diagnosing pneumonia and COVID-19 that rely on different characteristics. It is even more challenging when the target dataset has no labels available, i.e., unsupervised task transfer learning.

In this paper, we propose a novel \textbf{Task Adaptation Network~(TAN)} to solve this unsupervised task transfer problem. In addition to learning transferable features via domain-adversarial training, we propose a novel task semantic adaptor that uses the learning-to-learn strategy to adapt the task semantics. Experiments on three public COVID-19 datasets demonstrate that our proposed method achieves superior performance. Especially on COVID-DA dataset, TAN significantly increases the recall and F1 score by $5.0\%$ and $7.8\%$ compared to recently strong baselines. Moreover, we show that TAN also achieves superior performance on several public domain adaptation benchmarks.
\end{abstract}

\section{Introduction}
\label{sec-intro}

The COVID-19 pandemic is greatly threatening global public health.
In the battle against COVID-19, one critical challenge is to diagnose patients among a large number of people and provide necessary medical treatment so as to prevent further spread of the virus.
Nowadays there is a growing trend to use the screening of chest radiography images (CRIs) such as X-ray images~\cite{wang2020covid} for automated computer-aid diagnosis.

The diagnosis of COVID-19 based on chest X-ray images is a standard image classification problem with two classes: the infected and disinfected ones.
While deep neural networks (DNNs) have achieved great success for image classification, they often require a large amount of labeled images for training. Unfortunately, for \covid, large-scale annotations are costly and time-consuming to collect.
Therefore, a straightforward approach is to leverage transfer learning (TL) techniques~\cite{neyshabur2020being, yosinski2014transferable, pan2010survey,zamir2018taskonomy} to transfer knowledge from existing (abundant) typical pneumonia datasets (i.e., the \emph{source} domain) to COVID-19 (i.e., the \emph{target} domain) to facilitate the model learning.
In this paper, we mainly focus on the most challenging transfer setting where (1) the target domain has no labels and (2) the labels in the source and target domains are of different semantic meanings. We call this setting \emph{unsupervised task transfer}.

\begin{figure}[t!]
	\centering
	\subfigure[Our method performs transfer learning from typical pneumonia to \covid via feature distribution and task semantic adaptation.]{
		\centering
		\includegraphics[width=0.45\textwidth]{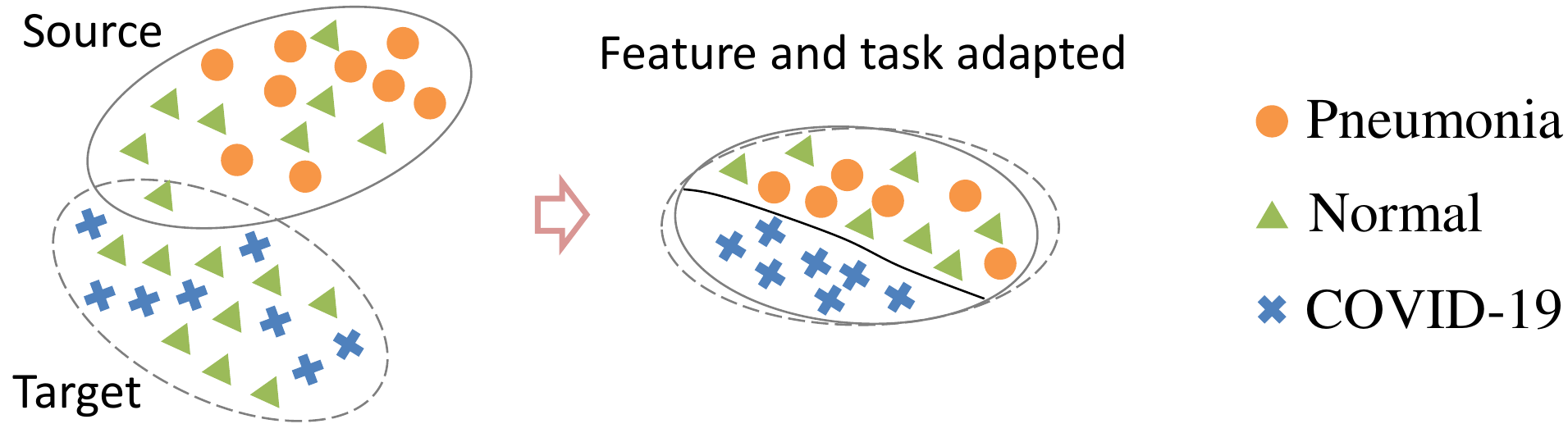}
		\label{fig-motiv-1}}
	\subfigure[Our method gives correct predictions by adapting both feature distributions and task semantics. Attention map shows that our method can capture the critical factors~\cite{wang2020covid} in the image that help detection.]{
		\centering
		\includegraphics[width=0.45\textwidth]{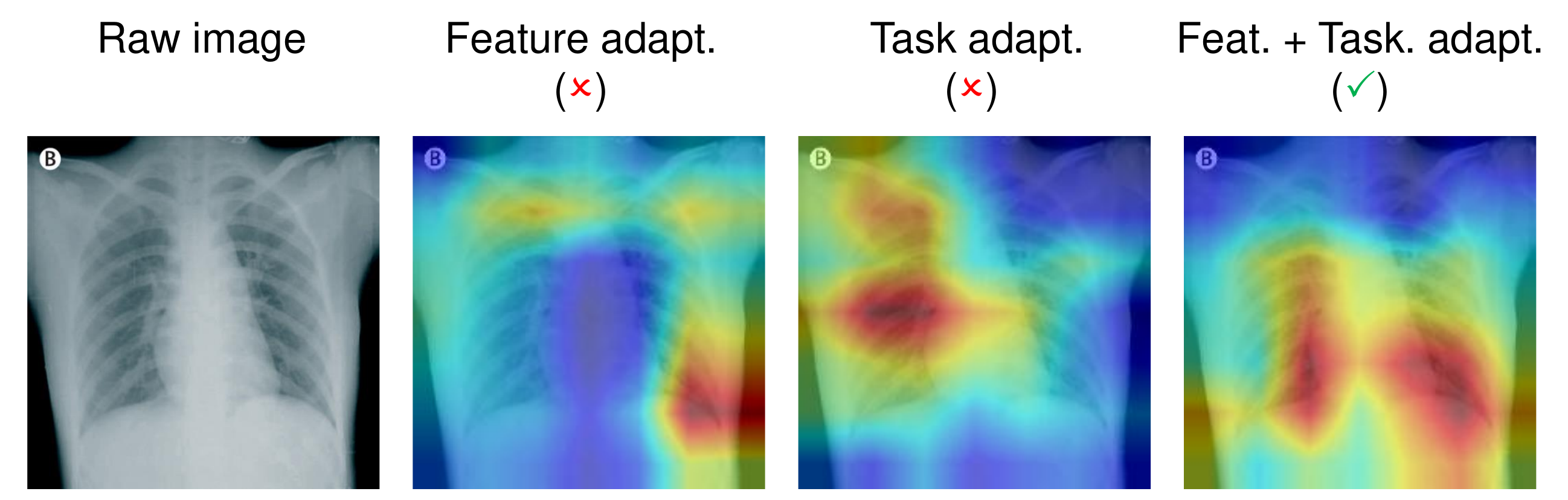}
		\label{fig-motiv-vis-onlyus}}
	\caption{Main idea and results visualization of our proposed TAN method. By adapting both the feature distribution and task semantics, TAN can eventually achieve accurate predictions by finding the most important critical factors.}

	\label{fig-motiv}
\end{figure}

In this unsupervised setting, the standard pretrain-finetune transfer paradigm becomes inapplicable, as there are no labeled images available in the target domain for finetuning.
This requires us to conduct unsupervised adaptation between two different tasks, i.e., train a model on the labeled source domain and adapt it to the target domain in an unsupervised manner.
Unsupervised adaptation presents two critical challenges that can result in \emph{negative transfer}~\cite{pan2010survey} and produces even worse performance than no transfer.
The first challenge is the \emph{feature distribution divergence}, which naturally exists since the distribution of these features differs from the source to the target domain.
Hence, feature distribution adaptation is necessary.
The second challenge is the \emph{task semantic difference} since diagnosing pneumonia and \covid are two related but different tasks that have different preferences on the critical factors~\cite{wang2020covid}.
Thus, the task semantics should also be adapted to maximize the transfer performance.
While existing domain adaptation (DA) methods are able to adapt feature distributions when the source and target tasks are identical (\emph{e.g.,} both of the source and target domains are classifying monitors under different background), they are not applicable to our problem~\cite{cui2020towards,zhang2019bridging, long2015learning, bousmalis2017unsupervised,tzeng2017adversarial}.

In this paper, we propose a \textbf{\methodfull~(\method)} for this unsupervised task transfer problem.
The concept of our method is illustrated in \figurename~\ref{fig-motiv-1}.
\method is able to learn transferable features across domains and tasks.
Concretely speaking, \method~firstly adopts domain-adversarial training to reduce the feature distribution divergence between domains.
However, only adapting features is not sufficient due to the task semantic difference.
\method~devises a novel permutation-invariant task semantic adaptor that uses the learning-to-learn strategy to handle task semantic difference.
We design a feature-critic training algorithm that effectively adapts task semantics using the pivot data.
\figurename~\ref{fig-motiv-vis-onlyus} gives an example of the activation map of \method to show its effectiveness in finding the critical factors~\cite{wang2020covid} by adapting both feature distribution and task semantics.

To sum up, this paper makes the following contributions:
\begin{enumerate}
	\item We propose a novel \methodfull~(\method) for unsupervised task transfer that addresses the feature distribution divergence and task semantic difference. Especially for the challenging task semantic adaptation, we propose a novel task semantic adaptor that leverages the learning-to-learning strategy to adapt cross-domain tasks.
	\item Experiments on three public COVID-19 chest X-ray image classification datasets demonstrate that \method outperforms several state-of-the-art baselines. To be more specific, on the challenging COVID-DA dataset, \method significantly improves the F1 score and recall by $7.8\%$ and $5.0\%$ respectively, compared to the second best baseline.
	\item Moreover, \method is a general and flexible method that also achieves superior performance on several public domain adaptation benchmarks including ImageCLEF-DA, Office-Home, and VisDA-2017.
\end{enumerate}

\section{Related Work}

Transfer learning (TL)~\cite{pan2010survey} is a useful technology to transfer the knowledge from existing source domains to the target domain, especially when the target domain has sparse or no labels.
Such label scarcity problem can be solved using TL by firstly pretraining on a large dataset such as ImageNet~\cite{deng2009imagenet} and then finetune the pretrained model on downstream tasks. This strategy is widely used in modern computer vision research~\cite{neyshabur2020being, sun2019meta,zamir2018taskonomy,yosinski2014transferable, donahue2014decaf}.
In a semi-supervised setting where the target domain has labels, Luo \textit{et al.}~\cite{luo2017label} proposed a domain and task transfer network to handle the different tasks using task semantic transfer.
However, when the target domain has no labels, the pretrain-finetue paradigm is not available.

When two tasks are related, multi-task learning (MTL)~\cite{caruana1997multitask,liu2019end,sener2018multi} can be used to learn transferable features to enhance their learning performance.
MTL also works when different tasks have labels available.
Meta-learning, or learning-to-learn~\cite{bengio1992optimization,finn2017model,santoro2016meta} aims to learn general knowledge from a bunch of tasks and then transfer to unseen tasks.
Meta-learning often works under the few-shot setting where different tasks have several labels available and does not explicitly reduce the feature distribution divergence between domains and tasks.
Zero-shot learning (ZSL)~\cite{pal2019zero,socher2013zero} focuses on classifying all unseen classes which can be seen as a general case to our problem that has one unseen category. In contrast, ZSL does not reduce the distribution divergence across domains.

Domain adaptation (DA) is a specific area of transfer learning~\cite{pan2010survey}.
DA aims at building cross-domain models by reducing the distribution divergence of a representation via some divergence measured by such as Maximum Mean Discrepancy (MMD)~\cite{gretton2012kernel}, KL or JS divergence, cosine similarity, and higher-order moments~\cite{zellinger2017central,long2015learning,tzeng2014deep,sun2016deep,zhang2018collaborative,wang2018visual}.
Another line of work relies on the generative adversarial nets~\cite{goodfellow2014generative} to learn domain-invariant features~\cite{zhang2019bridging,saito2018maximum, tzeng2017adversarial,ganin2014unsupervised}.
While great progress has been made, directly applying DA to our problem is not sufficient since DA generally works when two domains have identical categories.
The setting where two domains have different but overlapped tasks is also explored in recent open set DA~\cite{panareda2017open, saito2018open} and partial DA~\cite{zhang2018importance} methods. However, their purpose is to recognize the overlapped (common) categories rather than the unshared classes in the target domain.

\section{Methodology}

We introduce an unsupervised learning model which transfers information from a large labeled source domain, \source, 
to a target domain, \target, across different tasks. 
The goal being to learn a strong transferable target classifier $h: \mathcal{X}^{\mathcal{T}} \rightarrow \mathcal{Y}^{\mathcal{T}}$ 
that reduces both the feature distribution divergence and task semantic difference.

We assume the source domain contains $n^{s}$ images, $\vecx^s \in \mathcal{X}^{\mathcal{S}}$, with associated labels, 
$\vecy^s \in \mathcal{Y}^{\mathcal{S}}$. The target domain consists of $n^{t}$ unlabeled images, 
$\vecx^t \in \mathcal{X}^{\mathcal{T}}$.
\source and \target share the same feature space, i.e. $\mathcal{X}^{\mathcal{S}}=\mathcal{X}^{\mathcal{T}}$.


Unlike traditional domain adaptation approaches that assume a cross-domain distribution shift under a shared label space 
($\mathcal{Y}^{\mathcal{S}} = \mathcal{Y}^{\mathcal{T}}$), we aim to adapt both the feature distribution and task semantics, i.e., we consider the case where the tasks corresponding to source and target spaces are only \emph{similar but not identical},
$\mathcal{Y}^\mathcal{S} \ne \mathcal{Y}^\mathcal{T}$, like for the pneumonia and COVID-19. 
Even if these two classification problems can be treated as classifying images into $\{0,1\}$, their semantics are still different.\footnote{Although we can formulate both binary classification problems using $\mathcal{Y} = \{0,1\}$ for each, the semantic of label ``$1$'', i.e. the ``true label'', differs.}

\subsection{Overview}
\label{sec-method-overview}

\begin{figure}[t!]
	\centering
	\includegraphics[width=.48\textwidth]{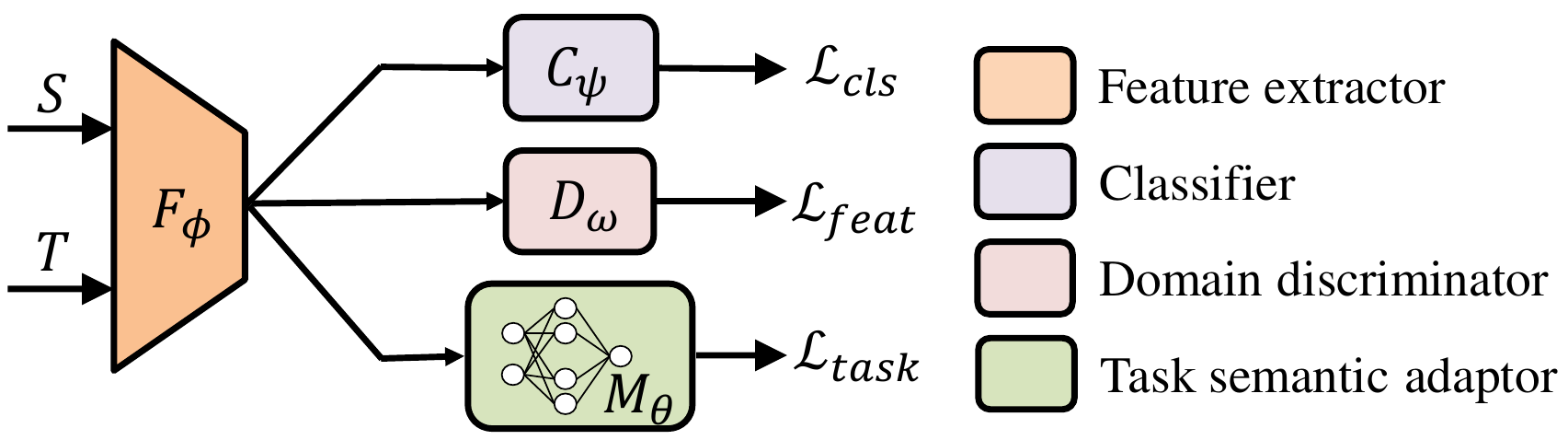}
	\caption{The architecture of the proposed \method method that consists of four modules: feature extractor $F_\phi$, classifier $C_\psi$, domain discriminator $D_\omega$ for feature distribution adaptation, and task semantic adaptor $M_\theta$ for task semantic adaptation.}
	\label{fig-framework}
	\vspace{-.1in}
\end{figure}

In this paper, we propose a novel \methodfull~(\method) to  adapt the feature distribution and task semantics.
We depict our overall model in \figurename~\ref{fig-framework}.
\method consists of four modules: feature extractor $F_\phi$, classifier $C_\psi$, domain classifier $D_\omega$, and task semantic adaptor $M_\theta$ with $\phi$, $\psi$, $\omega$, and $\theta$ as the learnable parameters.
Taking as inputs the labeled source examples, \method learns a latent feature space with $F_\phi$ and the binary classification network $C_\psi$ by standard supervised learning. Then, by also taking as inputs the unlabeled target domain, \method adapts the feature distributions and task semantics with the domain classifier $D_\omega$ and task semantic adaptor $M_\theta$, respectively.

Our model jointly optimizes over a source classification loss $\mathcal{L}_{cls}$, a feature distribution adaptation loss $\mathcal{L}_{feat}$,
and a task semantic transfer objective $\mathcal{L}_{task}$. Thus, the total objective of \method can be formulated as follows:
\begin{equation}
	\label{eq-ltn-all}
	\mathcal{L} = \mathcal{L}_{cls} + \lambda \mathcal{L}_{feat} + \mu \mathcal{L}_{task},
\end{equation}
where the hyperparameters $\lambda$ and $\mu$ determine the influence of the feature distribution adaptation and task semantic adaptation, respectively.
We use the cross-entropy loss $\ell^{(\text{CE})}$ to measure the classification error $\mathcal{L}_{cls}$ on the labeled source domain:
\begin{equation}
	\label{eq-ltn-cls}
	\mathcal{L}_{cls} = \mathbb{E}_{(\vecx, y) \in \mathcal{S}}~ \ell^{(\text{CE})}(C_{\psi}(F_{\phi}(\vecx)), y),
\end{equation}
where $\mathbb{E}$ denotes the expectation.
In the following sections, we will elaborate on the feature distribution adaptation and task semantic transfer modules.

\subsection{Feature distribution adaptation}
\label{sec-method-feat}

Considering the similarity between different tasks in \source and \target, the feature distribution adaptation module aims to reduce the cross-domain feature distribution divergence.
Inspired by the well-established domain-adversarial training~\cite{ganin2014unsupervised,ganin2016domain}, \method~designs a domain discriminator $D_\omega$ to adapt the cross-domain feature distributions.
Domain-adversarial training is a two-player game where the domain discriminator $D_\omega$ is trained to distinguish the source domain from the target domain, while the feature extractor $F_\phi$ tries to confuse the domain discriminator by learning domain-invariant features. 
These two players are trained adversarially, i.e., $\omega$ is trained by maximizing the loss on domain discriminator $D_\omega$ while $\phi, \omega$ are trained to minimize the loss of the feature extractor which is computed by the domain classifier.
This optimization procedure eventually minimizes the difference between the feature distributions on the two domains, measured by the Jensen-Shannon divergence~\cite{goodfellow2014generative}.
The adversarial training loss for feature distribution adaptation can be formulated as:
\begin{equation}
	\label{eq-ltn-adv}
	\begin{split}
		\mathcal{L}_{feat} = &~\mathbb{E}_{\vecx^s \in \mathcal{S}} \log [D_\omega (F_\phi (\vecx^s))]\\
		& + \mathbb{E}_{\vecx^t \in \mathcal{T}} \log [1 - D_\omega (F_\phi (\vecx^t))].
	\end{split}
\end{equation}

\subsection{Task semantic adaptation}
\label{sec-method-meta}

Domain-adversarial training is insufficient for our problem since it learns domain-invariant features across domains regardless of the tasks.
In our problem, even if the source and target tasks are similar which makes domain-adversarial training reasonable, these tasks are still not identical.
Adapting feature distributions unnecessarily ensures the adaptation across different task semantics.
Since features are highly correlated with the tasks, the difference in task semantics would harm the feature adaptation and the performance of knowledge transfer.

On this problem, we propose a task semantic adaptor $M_\theta$ that employs a learning-to-learn strategy to adapt task semantics by learning from the domain-adversarial features.
Learning-to-learn, or meta-learning~\cite{bengio1992optimization,finn2017model,santoro2016meta} aims to effectively leverage the datasets and prior knowledge of a task ensemble in order to rapidly learn new tasks often with a small amount of data.
Therefore, since the task semantics are difficult to model, we turn to using the learning-to-learning strategy.
The key idea is to let $M_\theta$ learn the adaptation ability from domain-adversarial training and then such ability can be utilized for task adaptation.
Therefore, when domain-adversarial training gradually encourages the features to be domain-invariant, the task semantic adaptor $M_\theta$ can gradually learn this ability using the learning-to-learn strategy and eventually also enforces the features to be task-invariant.
Hence, the task semantics can be adapted.

Technically speaking, the task semantic adaptor $M_\theta$ is implemented as an MLP network that can theoretically approximate any continuous functions~\cite{csaji2001approximation} to enable the superior adaptation. 
Denoting $\mathbf{F}_\phi^s$, $\mathbf{F}^t_\phi$ as the source- and target-domain features extracted by $F_\phi$, the task semantic adaptation loss $\mathcal{L}_{task}$ can be formulated as:
\begin{equation}
	\label{eq-task}
	\mathcal{L}_{task} = M_\theta (\mathbf{F}^s_\phi, \mathbf{F}^t_\phi).
\end{equation}

Unfortunately, it remains challenging to optimize the above equation w.r.t. $\theta$ for three reasons. 
Firstly, what property should $M_\theta$ satisfy to ensure that the task semantics can be adapted?
Secondly, how to maximally utilize the domain-invariant representations learned by adversarial training to better couple with the feature extractor for more effective training?
Thirdly, how to update the adaptor network parameters in training?

\paragraph{Permutation invariance.}
The task semantic adaptation is supposed to reduce the difference between two tasks, so it should be invariant to permutations of samples that represent each task distribution. 
Therefore, we design $M_\theta$ to be \emph{permutation-invariant} to the rows of its inputs, i.e., it should make no difference for different sample indices like the cases $[1, 2, 3]$ and $[3, 2, 1]$.
To enforce this property, we let $M_\theta$ take as inputs the pairwise distance between each element of $\mathbf{F}_\phi^s$ and $\mathbf{F}^t_\phi$, which is permutation-invariant~\cite{li2019feature}. Then, the task semantic loss can be represented as:
\begin{equation}
	\label{eq-task-inva}
	M_\theta (\mathbf{F}^s_\phi, \mathbf{F}^t_\phi) = \mathrm{MLP} (\flatten(\gram(\mathbf{F}_\phi^s, \mathbf{F}^t_\phi))),
\end{equation}
where \gram denotes the Gram matrix computed by pair-wise distance, \flatten is a flatten operation, and $\mathrm{MLP}$ denotes a multi-layer perceptron.

\paragraph{Pivot data.}
In a supervised case where target domain has labels, the task adaptor $M_\theta$ can be learned easily.
However, it becomes challenging in this unsupervised setting.
To maximally learn the adaptation ability from domain-adversarial training, $M_\theta$ is updated on a \emph{pivot data} \pivot.
It is a selected subset of both the source and target domains to maximally utilize the domain-adversarial training.
The pivot data are the data that with high confidence scores during the learning process, so they can be representatives of the domain-adversarial training.
As features are getting more domain-invariant, the classification performance on the target domain is gradually better, i.e., the pseudo labels for the target domain is getting more confident.
This pseudo-label training is widely adopted by most transfer learning literature~\cite{zhang2019bridging,tzeng2017adversarial}.
Therefore, based on the assumption that the adaptation ability can be learned from the domain-adversarial training, the task semantic adaptation could be better learned if $M_\theta$ can directly learn from samples with the most confident pseudo labels, i.e., the pivot data.
The number of pivot data is important to our problem: less pivot data will bring more confidence and less generalization while more pivot data will do the opposite, which is empirically evaluated in later experiments.

To be more specific, the pivot data can be represented as:
\begin{equation}
    \begin{split}
	\mathcal{P} =  \{\mathcal{P}_{\mathcal{S}}^{(c)}, &   \mathcal{P}_{\mathcal{T}}^{(c)}\}_{c  \in  \mathcal{Y}}, \\
	\mathcal{P}_{\mathcal{S}}^{(c)} = \{(\vecx_{j}, y_j=c)\}^{m}_{j=1}& 
	\mathcal{P}_{\mathcal{T}}^{(c)} = \{(\vecx_{j}, \hat{y}_j=c)\}^{m}_{j=1},
\end{split}
\end{equation}
where the data pairs are sorted as $\{(\vecx_{j}, \hat{y}_{j})\}$ in decreasing order of prediction score.
$\hat{y}$ is the predicted (pseudo) label on the target domain and $c$ is the class index.
We select the top $m$ instances for each class with high prediction scores (softmax probability that does not need the target labels).
This selection is iterated in the whole learning process. For the source domain, we directly use its ground-truth labels. In total, we select $m \cdot |\mathcal{Y}|$ pivot data for each domain.

\paragraph{Feature-critic training.}
Different from classification, there is no supervision information for the target domain that makes it hard to update $M_\theta$.
In this paper, we propose a \emph{feature-critic training} strategy~\cite{li2019feature} to update the task adaptor.
For notation brevity, we pack $\Phi=\{\phi, \psi, \omega\}$ for parameters other than $\theta$.

Our feature-critic training is introduced as follows.
Let $\Phi(t)$ and $\Phi(t+1)$ denote parameter values in two consecutive learning steps $t$ and $t+1$, respectively.
Our key assumption is that as the pseudo labels on pivot data are getting more confident, if $\Phi(t+1)$ is better than $\Phi(t)$ for task adaptation, it should produce lower risks and better classification performance.
Therefore, a reasonable feature-critic metric $M_\theta$ should evaluate a lower value for $\Phi(t+1)$ than for $\Phi(t)$. 
We thus update the feature-critic metric $M_\theta$ by minimizing difference $\mathcal{L}_{val}$ computed by $\Phi(t)$ and $\Phi(t+1)$:
\begin{equation}
	\mathcal{L}_{val} = \mathbb{E}_{\vecx\sim\mathcal{P}}~ \sigma(M_\theta(\vecx;\Phi(t+1))-M_\theta(\vecx;\Phi(t))),
\end{equation}
where $\sigma(\cdot)$ is an activation function.

\begin{algorithm}[t!]
	\caption{Learning algorithm of \method}
	\label{algo}
	\textbf{Input}: Source domain $\mathcal{S} = \{(\vecx^{s}_{i},y^{s}_{i})\}^{n_s}_{i=1}$, target domain $\pmb{T} = \{\vecx^{t}_{j}\}^{n_t}_{j=1}$, learning rate $\alpha, \beta$.
	
	\textbf{Output}: $\{\Phi^\ast\}$.
	\begin{algorithmic}[1] 
		\STATE Initialize $\Phi(0)$ and $\theta(0)$.
		\WHILE{not done}
		\STATE Build an assist model with its parameter inherited from the main model.
		\STATE \textbf{For} mini-batch data $B_s, B_t$ in $\mathcal{S}, \mathcal{T}$ \textbf{do}
		\STATE ~~~~Update $\Phi$ by Eq.~(\ref{eq-phi-update}).
		\STATE \textbf{End For}
		\STATE Select the data with the highest prediction confidence from $\mathcal{T}$ to construct pivot data $\mathcal{P}$.
		\STATE Update $\theta$ by Eq.~(\ref{eq-theta-update}). 
		\ENDWHILE
		\STATE \textbf{return} $\{\Phi^\ast\}$
	\end{algorithmic}
\end{algorithm}

\subsection{Training and inference}
The training process consists of two steps: 1)~update $\Phi$ for the feature extractor, classification layer and domain discriminator and 2)~update $\theta$ for the task semantic adaptor.

\paragraph{Update $\Phi$.}
This step is to update $\Phi$ for classification and domain-adversarial training. To enforce the update of $\theta$ in the next step, we construct an assist model which is a copy of the main model by inheriting the same architecture and parameters ($F_\phi, C_\psi, D_{\omega}$) and use it for calculating the loss by Eq.~(\ref{eq-ltn-all}) on all the labeled source domain and unlabeled target domain data.
Note that for updating the domain discriminator $D_\omega$, we do not use the mini-max optimization process and instead follow \cite{ganin2014unsupervised} to use the Gradient Reversal Layer for computing efficiency. Therefore, $\omega$ can be updated together with $\phi$ and $\psi$ in a single back-propagation.

After getting the training loss, denote $\alpha$ the learning rate of the main model, then $\Phi$ can be updated by:
\begin{equation}
	\label{eq-phi-update}
	\Phi(t+1)=\Phi(t)-\alpha \nabla_\Phi (\mathcal{L}_{cls} + \lambda \mathcal{L}_{feat} + \mu \mathcal{L}_{task}) | _{\Phi(t)}.
\end{equation}

\paragraph{Update $\theta$.}
This step is to update $\theta$ for the task adaptor $M_\theta$ using feature-critic training on the pivot data $\mathcal{P}$.
Denote $\beta$ its learning rate, then, $\theta$ can be updated by taking derivative of $\mathcal{L}_{val}$ on $\theta$:
\begin{equation}
	\label{eq-theta-update}
	\theta(t+1) = \theta(t)-\beta \nabla_{\theta} \mathcal{L}_{val}(\theta;\Phi(t),\Phi(t+1)) | _{\theta(t)},
\end{equation}
where $\Phi(t)$ and $\Phi(t+1)$ are parameters of the assist and main model, respectively.

The above two steps are used iteratively since the pseudo labels of the pivot data can be more confident and all the losses can be minimized.
In our experiments, we observe that the network will converge in dozens of epochs.

\begin{figure}[h]
	\centering
	\includegraphics[width=.48\textwidth]{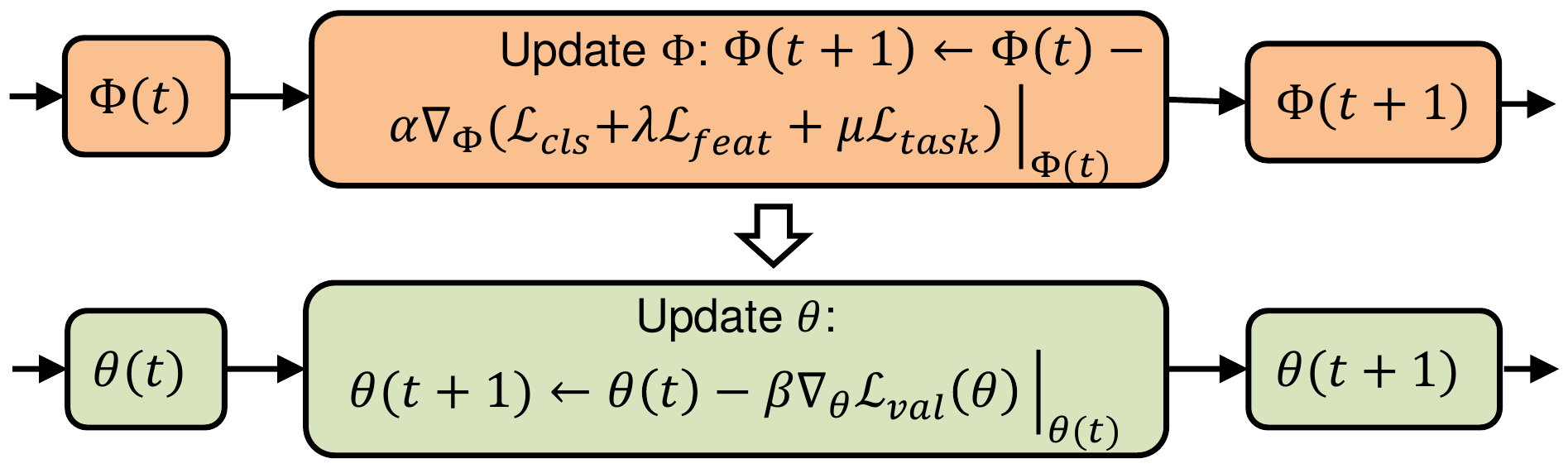}
	\caption{The learning process of \method is composed of two iterative steps: update $\Phi$ and update $\theta$.}
	\label{fig-learn-process}
\end{figure}

The training process of \method~is listed in Algorithm~\ref{algo} and \figurename~\ref{fig-learn-process}.
As for inference, we fix $\Phi$ to perform a single forward-pass to get the classification results for the test data.

\section{Experiments}

\subsection{Datasets and setup}

\begin{table}[t]
	\centering
	\caption{Statistical information of datasets}
	\label{tb-covid-dataset}
	\begin{tabular}{cccc}
		\toprule
		Dataset & \#normal & \#pneumonia & \#COVID-19 \\ \hline
		COVID-DA & 9,039 & 2,306 & 318 \\ 
		Bacterial & 3,270 & 3,001 & 1,281 \\ 
		Viral & 3,270 & 1,656 & 1,281 \\ \bottomrule
	\end{tabular}
	\vspace{-.1in}
\end{table}

We evaluate \method on public \covid chest X-ray datasets.
COVID-DA~\cite{zhang2020covidda} contains three categories: typical pneumonia, normal, and COVID-19.
Curated-COVID~\cite{curated-covid} is a larger dataset that consists of four categories: normal, bacteria pneumonia, viral pneumonia, and COVID-19.
Since our focus is on investigating the performance on transfer learning from pneumonia to COVID-19, we split Curated-COVID into two datasets where each one only contains one type of pneumonia and COVID-19.
We name them \emph{Bacterial} and \emph{Viral} accordingly.
Together we have three datasets as shown in \tablename~\ref{tb-covid-dataset}.

For COVID-DA, we follow \cite{zhang2020covidda} to construct the source, target, and validation domains.
For Bacterial and Viral datasets, we construct their source and target domains by taking all pneumonia (for source) and all COVID-19 (for target) samples accordingly.
As for the normal category, we split them evenly into two domains.
We further leave $20\%$ of the target domain for validation.
Eventually, there are two classes in the source domain: normal and pneumonia; and there are normal and COVID-19 classes in the target and validation datasets.
In this case, the source domain does not contain any COVID-19 samples, which makes our problem harder than traditional transfer learning and domain adaptation.
The detailed domain split information is presented in appendix.

\subsection{Comparison methods}

We compare the performance of \method with three categories of methods: (1) deep and traditional transfer learning baselines, (2) deep diagnostic methods, and (3) unsupervised DA methods.

The deep and traditional TL baselines include:
\emph{Pretrain-only}, which trains a network on the source domain, and then directly apply the pretrained model on the target domain.
\emph{Target-train}, which is an ideal state and only for comparison since there are no labels for the target domain. We directly use several extra labeled COVID-19 data from the dataset (they are $30\%$ of the target domain data) and train a network on these data. Then, we apply prediction on the target data.
\emph{Pretrain-finetune}, which is a standard TL paradigm that finetunes the source pretrained model on the labeled target domain data. 
Note that \emph{Target-train} and \emph{Pretrain-finetune} methods use labeled data from the target domain, which is not available for \method and other baselines. They are just for a comparison with the ideal setting of machine learning, i.e. the conventional supervised learning.

We choose DLAD~\cite{zhang2020covid} as the deep diagnostic method. 
The unsupervised DA methods are DANN~\cite{ganin2014unsupervised}, MCD~\cite{saito2018maximum}, CDAN+TransNorm~\cite{wang2019transferable}, MDD~\cite{zhang2019bridging}, and BNM~\cite{cui2020towards}.
All methods are using ResNet-18~\cite{he2016deep} as the backbone network following~\cite{zhang2020covidda}.
The results of these methods are obtained from \cite{zhang2020covidda} to ensure a fair comparison. Note that we do not compare with \cite{zhang2020covidda} since it is a semi-supervised method that requires labeled data on the target domain.

For \method, we use the mini-batch SGD with Nesterov momentum of $0.9$ to optimize the main and the meta-network with batch size set as $16$.
The learning rate $\alpha$ of the main model changes by following~\cite{ganin2014unsupervised}: $\alpha_{k} = \frac{\alpha}{(1+\gamma k)^{-\upsilon}}$, where $k$ is the training iteration, $\gamma=0.001$, $\alpha=0.004$, and decay rate $\upsilon=0.75$.
The learning rate $\beta$ for $M_\theta$ is set to be $0.0005$.
$M_\theta$ uses a $\mathrm{in}-128-64-1$ MLP structure where $\mathrm{in}$ is the dimension of input matching features.
We grid search the value of $\lambda$ and $\mu$ in the range $[0.01, 0.05, 0.1, 0.5, 1, 5, 10]$ for the best performance.

During training, the labels of the target domain are not available and they are only be used for evaluation.
For this binary classification problem, we use F1, Precision (P), and Recall (R) as the evaluation metrics.
We do not use ROC/AUC since we are more interested in the recall and F1 in this specific disease diagnosis problem.
The results are the average accuracy of ten trials.

\begin{table}[t!]
	\centering
	\caption{\upshape Results on COVID-DA dataset (typical pneumonia $\rightarrow$ COVID-19, ResNet-18).}
	\label{tb-covid}
	
	\centering
		\begin{tabular}{lccc}
			\toprule
			Method & P (\%) & R (\%) & F1 (\%) \\ \hline
			Pretrain-only & 63.5 & 66.7 & 65.0 \\
			Target-train \textit{(ideal)} & \textbf{91.7} & 55.0 & 68.8 \\
			Pretrain-finetune \textit{(ideal)} & 56.3 & 75.0 & 64.3 \\
			DLAD~\cite{zhang2020covid} & 62.0 & 73.3 & 67.2 \\
			DANN~\cite{ganin2014unsupervised} & 61.4 & 71.7 & 66.2 \\
			MCD~\cite{saito2018maximum} & 63.2 & 60.0 & 61.5 \\
			CDAN+TransNorm~\cite{wang2019transferable} & 85.0 & 39.2 & 63.7 \\ 
			MDD~\cite{zhang2019bridging} & 74.0 & 60.0 & 67.0 \\
			BNM~\cite{cui2020towards} & 43.0 & 75.0 & 56.0 \\
			\method~(Ours) & 70.6 & \textbf{80.0} & \textbf{75.0}\\
			\bottomrule
		\end{tabular}
\end{table}

\subsection{Results and analysis}
The results on COVID-DA dataset are presented in \tablename~\ref{tb-covid}.
Here we use the 95\% confidence interval, where the corresponding value of z is 1.96. The computed
confidence interval r is around 1.3\%.
Note that we do not list the accuracy results since all methods achieve similarly high accuracy values in this binary classification problem.
On COVID-DA dataset, our proposed \method achieves a recall of $80.0\%$ and F1 of $75.0\%$, which significantly outperforms the second best baselines by $5.0\%$ in recall and $7.8\%$ in F1 score.
Pretrain-only and Finetune do not get good performance because of the feature distribution divergence and task semantic difference, which drastically limit their performance.
Compared to other DA methods (DANN, MCD, CDAN, MDD and BNM), \method also achieves better recall and F1 score.
While the precision of CDAN is better than ours, its recall and F1 is not competent.
Compared to the ideal state which is training on labeled target data, it is surprising to find that even working in fully unsupervised setting, our proposed \method~can achieve better recall and F1 score.

\begin{table}[t!]
	\centering
	\caption{Results on Bacterial dataset (bacterial pneumonia $\rightarrow$ \covid, ResNet-18)}
	\label{tb-result-bacteria}
	\begin{tabular}{cccc}
		\toprule
		Method & P (\%) & R (\%) & F1 (\%) \\ \hline
		Pretrain-only & 92.2 & 91.6 & 91.9 \\
		DAN~\cite{long2015learning} & 87.5 & 97.6 & 92.3 \\
		DANN~\cite{ganin2014unsupervised} & 93.1 & 96.8 & 94.9 \\
		MDD~\cite{zhang2019bridging} & 91.3 & 97.6 & 94.4 \\
		BNM~\cite{cui2020towards} & 90.6 & 98.9 & 94.6 \\
		\method (Ours) & 92.0 & \textbf{99.1} & \textbf{95.4}  \\
		\bottomrule
	\end{tabular}
	\vspace{-.1in}
\end{table}

\begin{table}[t!]
	\centering
	\caption{Results on Viral dataset (viral pneumonia $\rightarrow$ \covid, ResNet-18)}
	\label{tb-result-viral}
	\begin{tabular}{cccc}
		\toprule
		Method & P (\%) & R (\%) & F1 (\%) \\ \hline
		Pretrain-only & 85.4 & 65.9 & 74.4 \\
		DAN~\cite{long2015learning} & 77.4 & 62.7 & 72.2 \\
		DANN~\cite{ganin2014unsupervised} & 67.4 & 83.3 & 74.5 \\
		MDD~\cite{zhang2019bridging} & 92.6 & 94.6 & 90.6 \\ 
		BNM~\cite{cui2020towards} & 88.7 & 97.7 & 93.0 \\ 
		\method (Ours) & 91.5 & \textbf{99.2} & \textbf{95.2} \\ \bottomrule
	\end{tabular}
\end{table}

\begin{figure*}[t!]
	\centering
	\subfigure[All methods give correct predictions]{
		\centering
		\includegraphics[width=0.48\textwidth]{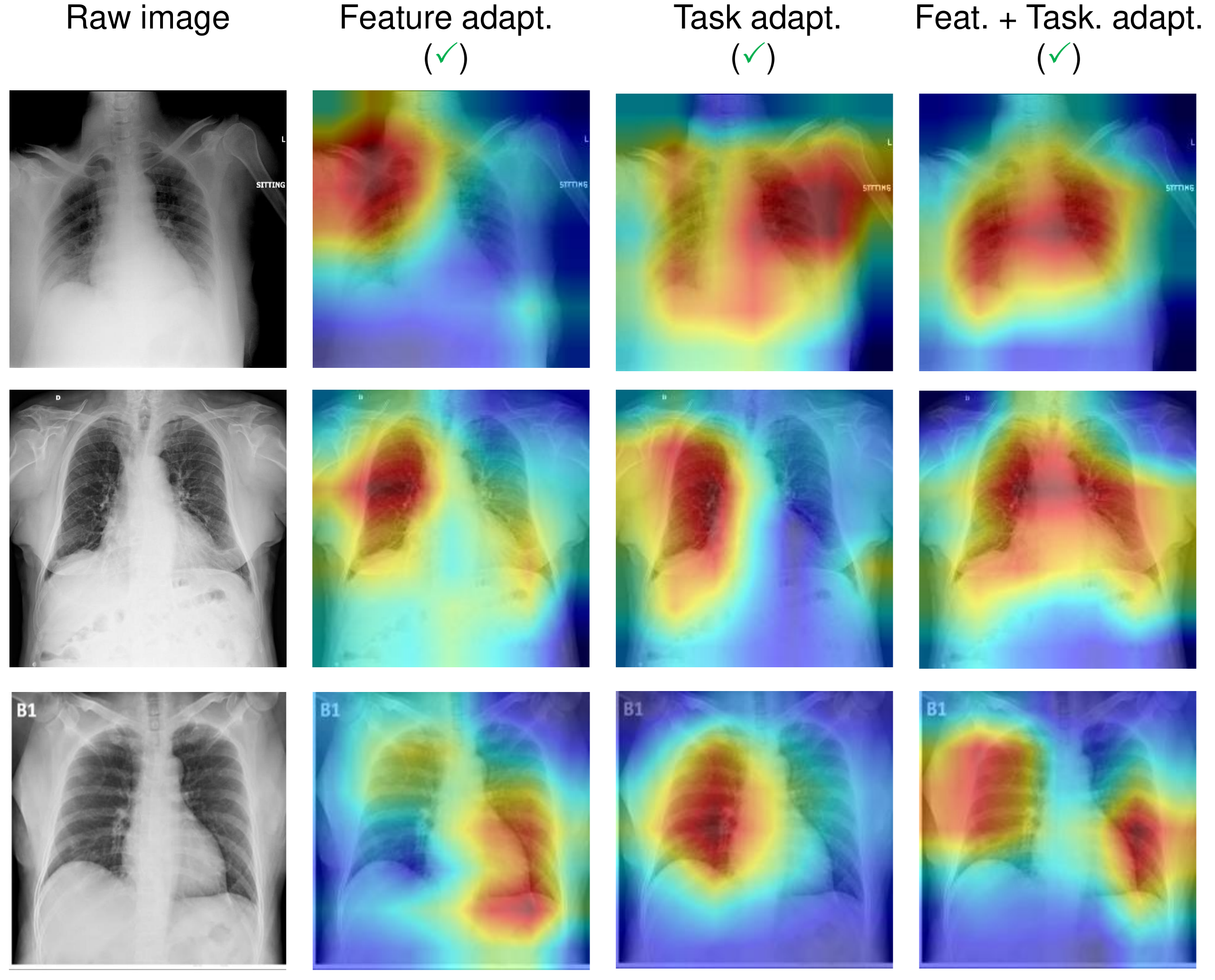}
		\label{fig-vis-allcorrect}}
	\subfigure[Adapting both feature and tasks give correct predictions]{
		\centering
		\includegraphics[width=0.48\textwidth]{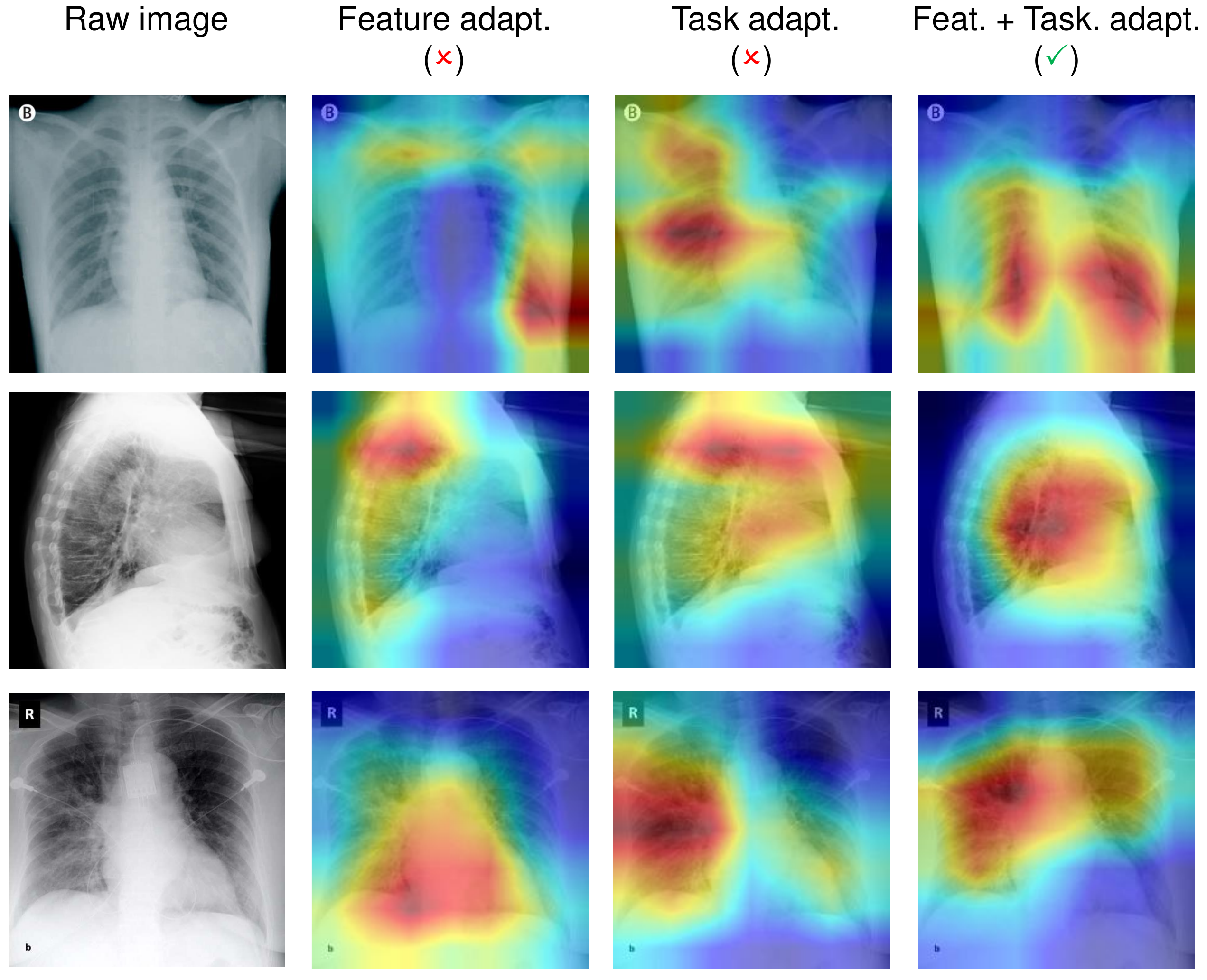}
		\label{fig-vis-onlyus}}
	
	\caption{Results visualization using Grad-CAM~\cite{selvaraju2017grad} to show the attention weight of different adaptation modules. In some easy cases in (a), either adapting feature distributions or task semantics alone can achieve correct predictions. However for some hard cases in (b), it requires to adapt both feature distributions and task semantics to get correct predictions.}
	\vspace{-.1in}
	\label{fig-vis-attention}
\end{figure*}

The results on Bacteria and Viral datasets are shown in \tablename~\ref{tb-result-bacteria} and \tablename~\ref{tb-result-viral}, respectively.
In these two datasets, we do not compare with the two ideal states target-train and pretrain-finetune since their performances are consistently better due to increased \covid samples in these datasets.
Our \method still significantly outperforms other baselines in these large datasets.
Same as on the COVID-DA dataset, pretrain-only gives the worst results due to feature distribution divergence and task semantic difference.
While domain adaptation methods (DAN, DANN, MDD and BNM) outperform pretrain-only by aligning the feature distributions, they do not adapt the task semantics.

Comparing the results from all three tables, we see that with the numbers of unsupervised target domain samples increase, the transfer learning performance tends to become better, i.e., the results in \tablename~\ref{tb-result-bacteria} and \tablename~\ref{tb-result-viral} are generally better than the results in \tablename~\ref{tb-covid}.
This is because more representative knowledge can be learned when domains have more samples that facilitates transfer learning.
In all situations, \method consistently achieves the best performance.

\subsection{Ablation study}
To further evaluate the effectiveness of \method, we conduct an ablation study in \tablename~\ref{tb-ablation-covid}.
For task adaptation, we remove the domain-adversarial training module and directly let the model learn from the classification loss.
For feature adaptation, we simply remove the task adaptation module.
The results show the following observations.
Firstly, the classification loss alone does not get good results, indicating the existence of feature distribution divergence and task semantic difference.
Secondly, better performance can be achieved by combining the feature distribution adaptation and task semantic adaptation modules, indicating that both adaptation modules are effective.
Thirdly, the best performance is achieved by combining both of the feature adaptation and task semantic adaptation modules, which proves that both of them are important in this problem.

\begin{table}[h]
	\centering
	\caption{Ablation study on \method}
	\label{tb-ablation-covid}
	\begin{tabular}{cccc}
		\toprule
		Variants & P (\%) & R (\%) & F1 (\%) \\ \hline
		$\mathcal{L}_{cls}$ & 63.5 & 66.7 & 65.0 \\ 
		$\mathcal{L}_{cls} + \mathcal{L}_{task}$ & 75.0 & 65.0 & 69.6 \\ 
		$\mathcal{L}_{cls} + \mathcal{L}_{feat}$ & 61.4 & 71.7 & 66.2 \\ 
		$\mathcal{L}_{cls} + \mathcal{L}_{feat} + \mathcal{L}_{task}$ & 70.6 & 80.0 & 75.0 \\ 
		\bottomrule
	\end{tabular}
	\vspace{-.1in}
\end{table}

\subsection{Visualization study}
The change of lung is critical factor for diagnosing \covid, which could be visualized to study the effectiveness of our method.
Therefore, we visualize the attention maps for several \covid images using the Gradient-weighted Class Activation Mapping (Grad-CAM)~\cite{selvaraju2017grad} in \figurename~\ref{fig-vis-attention}.
The shadow area in the figures is the lung area and the heat map denotes the activation weights for the model.
Specifically, \figurename~\ref{fig-vis-allcorrect} shows the cases where all adaptation modules give correct predictions, while \figurename~\ref{fig-vis-onlyus} shows the cases where wrong predictions are given by only feature distribution adaptation and only task semantic adaptation, and correct predictions are given by adapting both the feature distributions and task semantics.

\begin{figure*}[t!]
	\centering
	\subfigure[Pivot data]{
		\centering
		\includegraphics[width=0.23\textwidth]{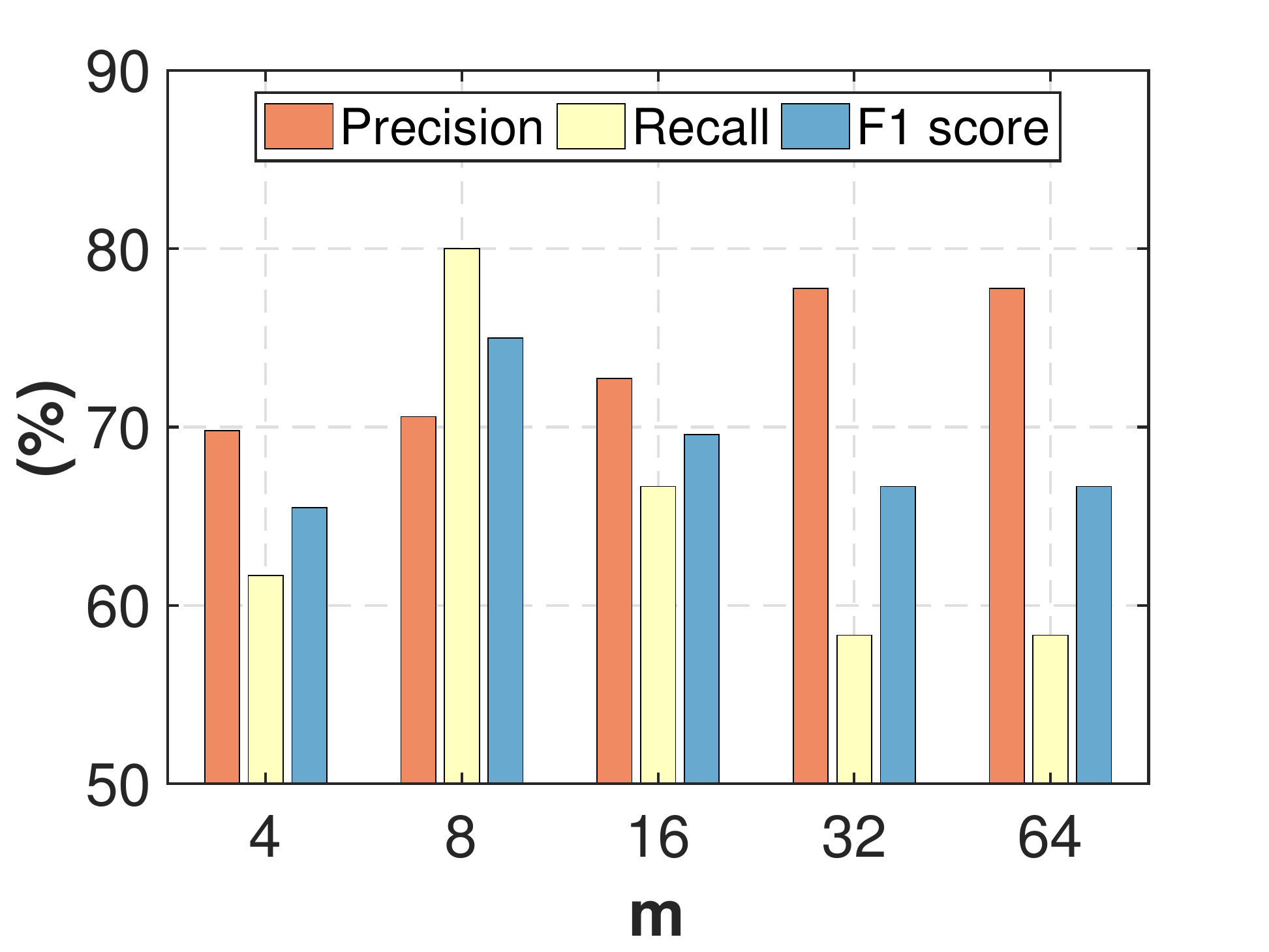}
		\label{fig-meta}} \,
	\subfigure[Permutation invariance]{
		\centering
		\includegraphics[width=0.23\textwidth]{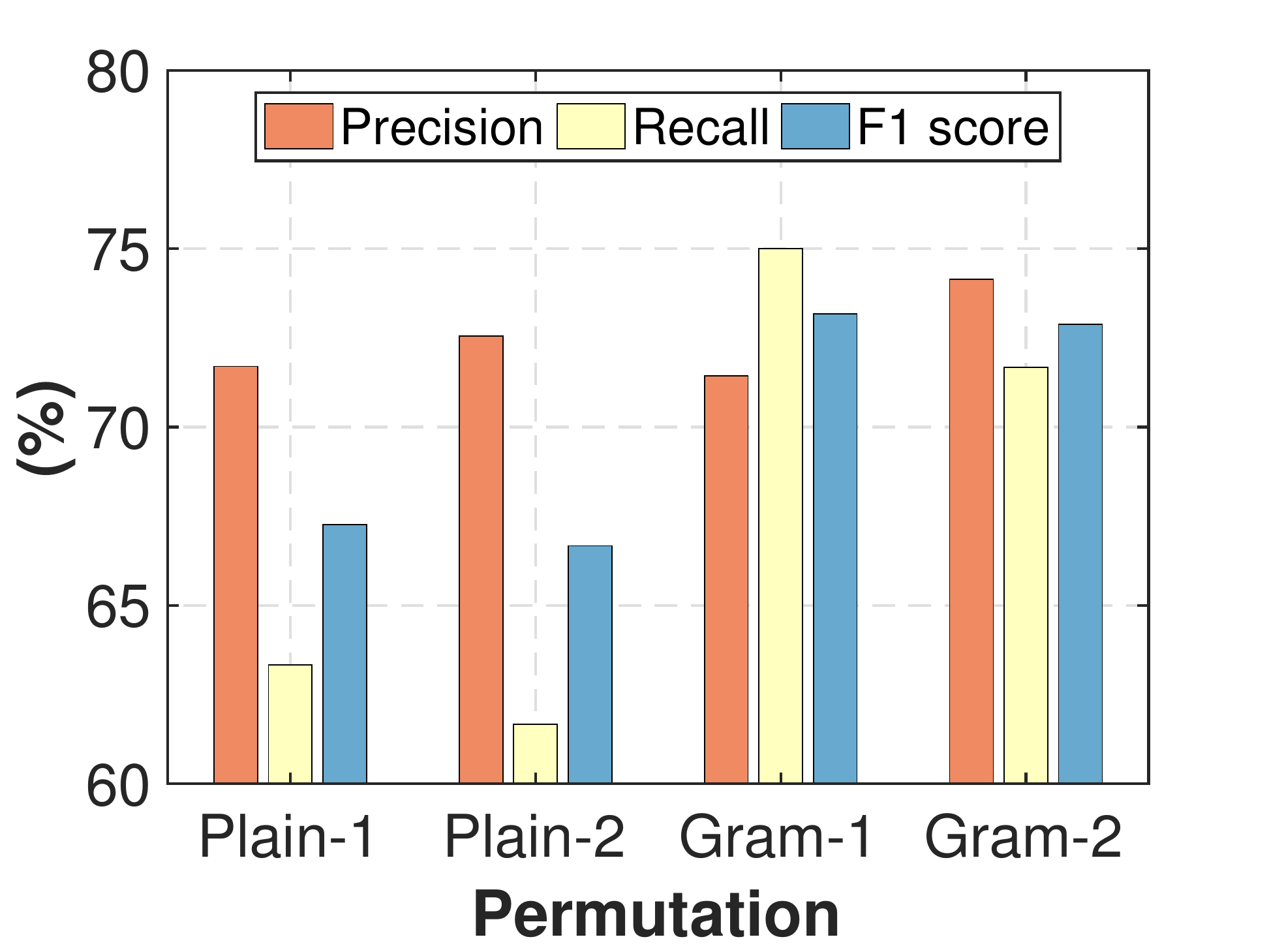}
		\label{fig-permutation}}
	\subfigure[Distribution distance]{
		\centering
		\includegraphics[width=0.23\textwidth]{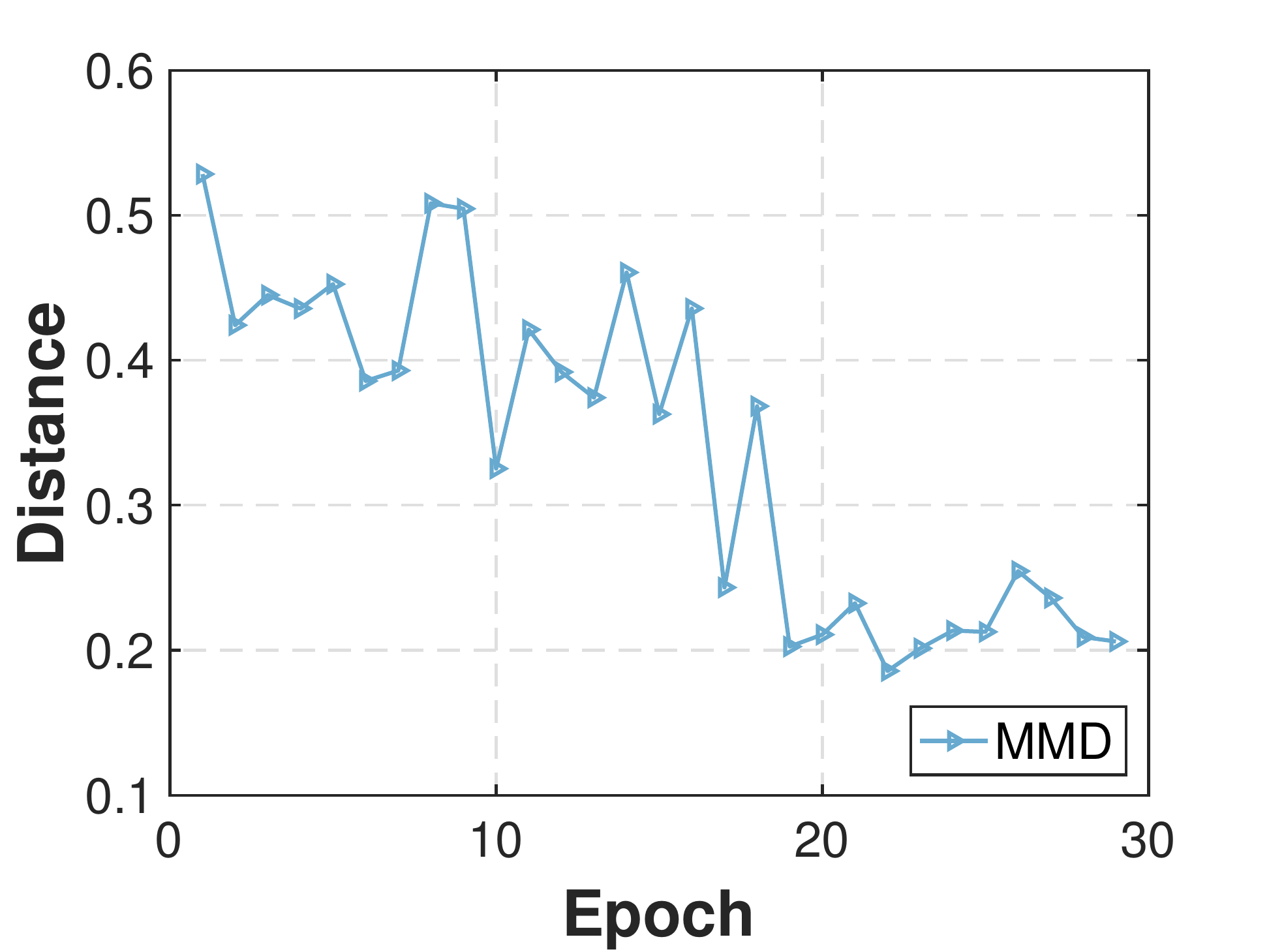}
		\label{fig-distribution}} \,
	\subfigure[Convergence]{
		\centering
		\includegraphics[width=0.23\textwidth]{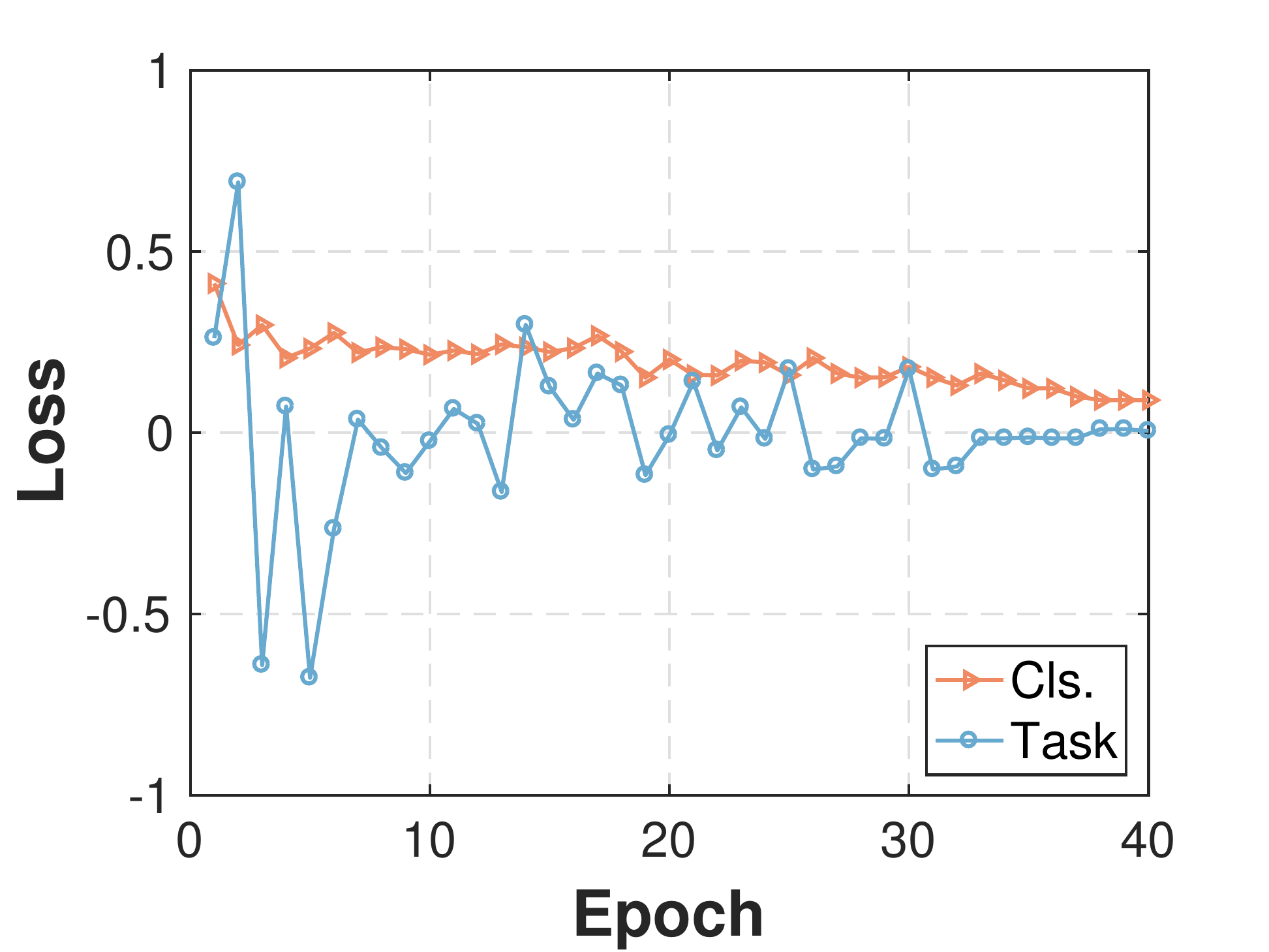}
		\label{fig-conv}}
	
	\caption{Detailed analysis of our method. (a) The number of pivot data in the target domain. (b) Evaluation of the permutation-invariant property of task semantic adaptor. (c) Distribution distance of feature adaptation. (d) Convergence analysis.}
	\label{fig-analysis}
	\vspace{-.1in}
\end{figure*}

From these results, we observe that in general cases where the samples are easy to classify, all adaptation methods give correct prediction.
However, when the samples are hard such as the second one in \figurename~\ref{fig-vis-onlyus} which is sideways, it becomes harder for the model to classify.
In this case, only adapting feature distributions or task semantics are not sufficient.
Our proposed \method~can perform reasonably well in all situations.

\subsection{Further analysis}

\paragraph{Pivot data}
We empirically analyze size $m$ of the pivot data $\mathcal{P}$.
It is obvious that a larger $m$ will bring more uncertainty, and a smaller $m$ is likely to make the meta-network unstable.
We record the performance of \method using different values of $m$ on COVID-DA dataset in \figurename~\ref{fig-meta}.
The results indicate that \method is robust to $m$ and a small $m$ can lead to competitive performance.
Therefore, we set $m=8$ in experiments for computational efficiency.
We also compare different pivot data selection strategies in the appendix.

\paragraph{Permutation invariance}
We evaluate the permutation invariance property of $M_\theta$.
\figurename~\ref{fig-permutation} shows the two random results of using Gram matrix as inputs (i.e., permutation-invariant) and the raw features (i.e., the permutation-variant case, which we denote as `plain' in the figure).
While both networks outperform other baselines in \tablename~\ref{tb-covid}, the Gram matrix gets the best performance, indicating that permutation invariance is important in the task semantic adaptor.

\paragraph{Feature distribution}
We compute the feature distribution distance using the maximum mean discrepancy (MMD)~\cite{borgwardt2006integrating} as shown in \figurename~\ref{fig-distribution}.
The results indicate that our proposed \method can gradually reduce the distribution distance.

\paragraph{Convergence and parameter sensitivity analysis}
We record the loss for classification and task semantic adaptation in \figurename~\ref{fig-conv}.
The results show that although \method~involves both the classification and task adaptation networks, it can quickly reach a steady performance.
This makes it easy to train in real applications.
We also empirically analyze the sensitivity of the two trade-off parameters $\lambda$ and $\mu$ and present the results in appendix, which shows that our method is relatively robust to these parameters.

\section{Experiments on other datasets}

Although our main focus is \covid, \method is not limited to this problem.
In fact, \method can be applied to any dataset with similar setting and standard DA datasets.

\paragraph{Constructed dataset.}
We construct a new dataset from the VisDA-2017 DA challenge~\cite{visda2017} by selecting its two random classes.
In this dataset, the source domain is rendered 3D objects and the target domain is natural images.
We call this constructed dataset \emph{VisDA-binary}.
Specifically, the source domain contains two classes: {\ttfamily train} (16,000) and {\ttfamily truck} (9,600) and the target domain contains {\ttfamily train} (4,236) and {\ttfamily bus} (4,690). The goal is to maximize the binary classification performance on the target domain, especially on class {\ttfamily bus}.
This dataset is more balanced with more samples than COVID-19, which can be regarded as its complement.
The results in \tablename~\ref{tb-visda} show that \method outperforms all other comparison methods in recall and F1 score.

\begin{table}[t!]
	\centering
	\caption{Results on VisDA for binary classification (ResNet-50)}
	\label{tb-visda}
	\begin{tabular}{cccc}
		\toprule
		Method & P (\%) & R (\%) & F1 (\%) \\ \hline
		Pretrain-only & 64.6 & 71.9 & 68.0 \\ 
		DAN~\cite{long2015learning} & 78.4 & 54.1 & 64.0 \\ 
		DANN~\cite{ganin2014unsupervised} & 89.2 & 70.1 & 78.5 \\ 
		BNM~\cite{cui2020towards} & 79.6 & 60.5 & 68.8 \\ 
		\method~(Ours) & 75.3 & \textbf{88.0} & \textbf{81.2} \\ \bottomrule
	\end{tabular}
\end{table}

\paragraph{Standard DA banchmark.}

We also evaluate the performance of \method~on several standard domain adaptation benchmarks.
\tablename~\ref{tb-clef} shows the results of \method~against other strong baselines on the ImageCLEF-DA~\cite{imageclef} dataset.
We see that although \method is not specifically designed for traditional DA tasks, it still achieves competitive performance.
We show the results on Office-Home~\cite{venkateswara2017deep} dataset in appendix where \method also produces competitive performance.

\begin{table}[t!]
	\caption{\upshape Accuracy~(\%) on ImageCLEF-DA for UDA (ResNet-50).}
	\label{tb-clef}
	\centering 
	\resizebox{.48\textwidth}{!}{
		\begin{tabular}{lccccccc}
			\toprule
			Method  & I$\rightarrow$P & P$\rightarrow$I & I$\rightarrow$C & C$\rightarrow$I & C$\rightarrow$P & P$\rightarrow$C & AVG \\ \hline
			ResNet~\cite{he2016deep}  & 74.8 & 83.9 & 91.5 & 78.0 & 65.5 & 91.2 & 80.7 \\
			DAN~\cite{long2015learning}     & 75.0 & 86.2 & 93.3 & 84.1 & 69.8 & 91.3 & 83.3 \\
			DANN~\cite{ganin2014unsupervised}    & 75.0 & 86.0 & 96.2 & 87.0 & 74.3 & 91.5 & 85.0 \\
			D-CORAL~\cite{sun2016deep} & 76.9 & 88.5 & 93.6 & 86.8 & 74.0 & 91.6 & 85.2 \\
			CAN~\cite{zhang2018collaborative}     & 78.2 & 87.5 & 94.2 & 89.5 & 75.8 & 89.2 & 85.7 \\ 
			JAN~\cite{long2016deep}     & 76.8 & 88.0 & 94.7 & 89.5 & 74.2 & 91.7 & 85.8 \\
			MADA~\cite{pei2018multi}    & 75.0 & 87.9 & 96.0 & 88.8 & 75.2 & 92.2 & 85.8 \\
			CDAN~\cite{long2018conditional} & 77.7 & 90.7 & \textbf{97.7} & 91.3 & 74.2 & 94.3 & 87.7  \\
		    TransNorm~\cite{wang2019transferable} & 78.3 & 90.8 & 96.7 & \textbf{92.3} & 78.0 & 94.8 & 88.5 \\
			\method (Ours) & \textbf{78.7} & \textbf{91.0} & 97.0 & 92.0 & \textbf{79.7} & \textbf{96.0} & \textbf{89.1} \\ \bottomrule
		\end{tabular}
	}
	\vspace{-.1in}
\end{table}

\section{Conclusions and Future Work}

In this paper, we propose a \methodfull~(\method) for \covid chest X-ray image classification by transferring knowledge from the typical pneumonia.
\method~can adapt both of the cross-domain feature distributions and task semantics to produce accurate prediction on the target domain.
Specifically for task semantic adaptation which is hard to model, we design a semantic adaptor that leverages the learning-to-learn strategy to learn the adaptation ability from the domain-adversarial training.
Experiments on several public datasets show that \method significantly outperforms other comparison approaches.
Moreover, \method~can also achieve competitive performance on several domain adaptation benchmarks.

In the future, we plan to apply \method to more fine-grained \covid diagnosis tasks such as detection and segmentation. \method~can also be applied to other \covid data modalities like CT scans.
In addition, we also plan to apply \method~to other similar transfer learning problems.

{\small
	\bibliographystyle{ieee_fullname}
	\bibliography{cvpr21}
}

\newpage

\appendix

\input{cvpr21_supp}

\end{document}

%% file: cvpr21_supp.tex
\section{Evaluation of pivot data}
In our main experiments, we set the $m=8$ for the pivot data to select the top $m$ samples belonging to one class with the highest probabilities.
For pivot data construction, we further evaluate the performance of other two strategies: (1) select random $m$ samples for each class and (2) select the bottom $m$ samples for each class.

\begin{figure}[htbp]
	\centering
	\includegraphics[width=.3\textwidth]{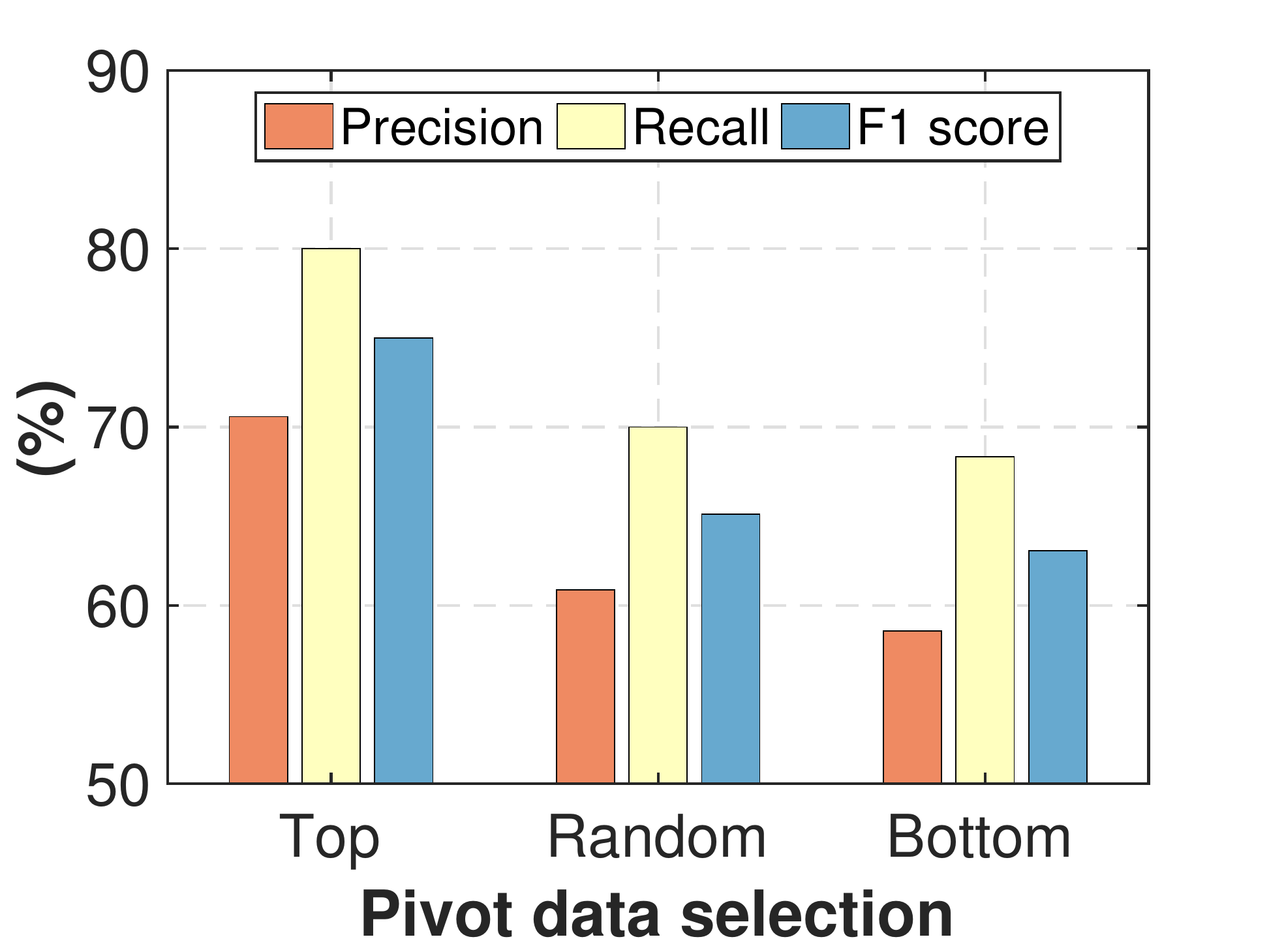}
	\caption{Different strategies on pivot data selection.}
	\label{fig-pivot}
\end{figure}

Here, `bottom $m$' is the opposite of top $m$ in the main paper, which is selecting the $m$ samples with the lowest probabilities.
The results in \figurename~\ref{fig-pivot} show that the top $m$ strategy achieves the best performance.

\section{ROC curve}
\begin{figure}[htbp]
	\centering
	\includegraphics[width=.3\textwidth]{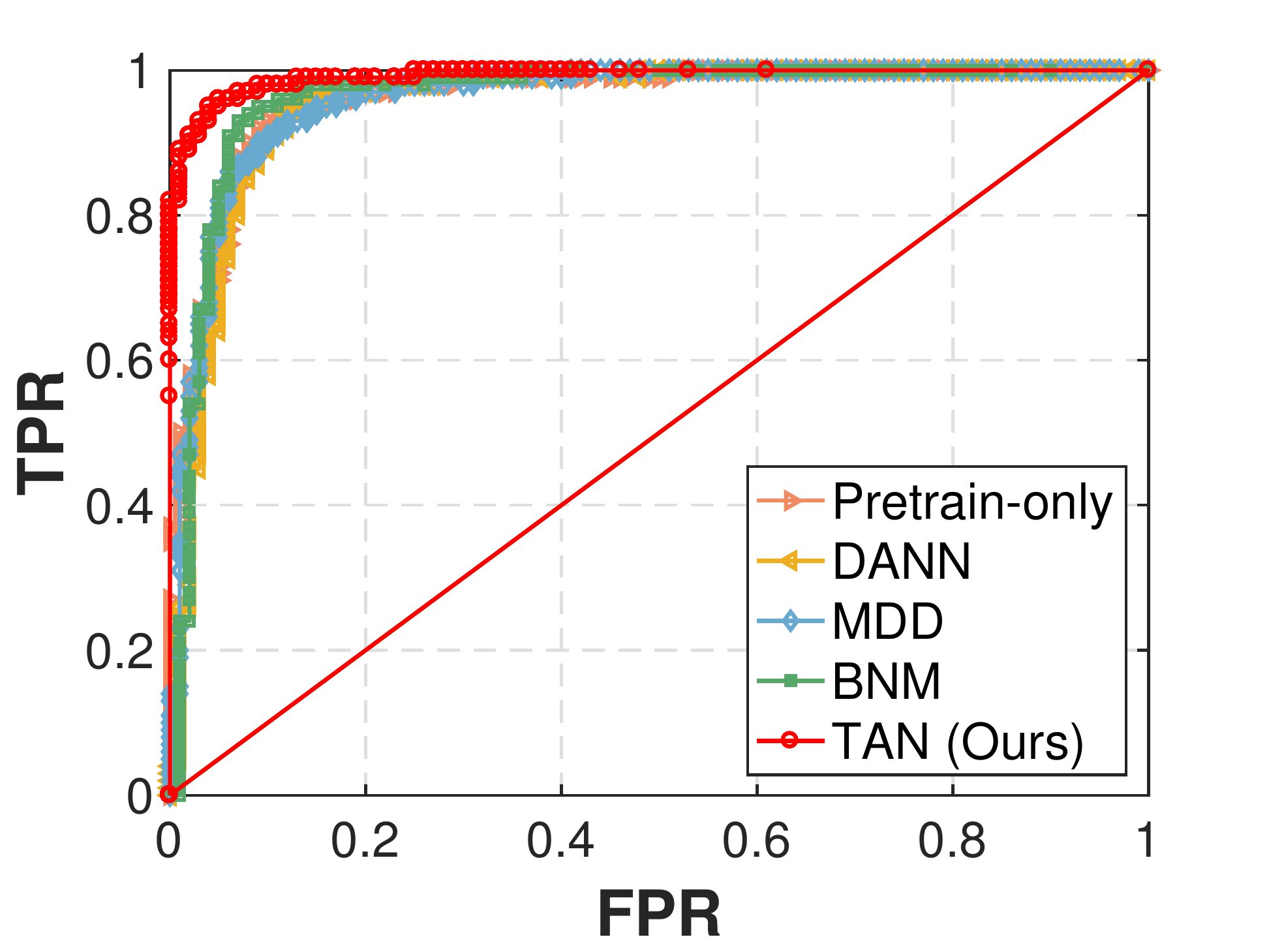}
	\caption{ROC curve.}
	\label{fig-roc}
\end{figure}

As for evaluation metrics, other than precision, recall, and F1 score, we further draw the ROC curve in \figurename~\ref{fig-roc}.
The results show that our proposed \method can achieve superior performance on this binary classification problem.

\begin{figure}[htbp]
	\centering
	\includegraphics[width=.4\textwidth]{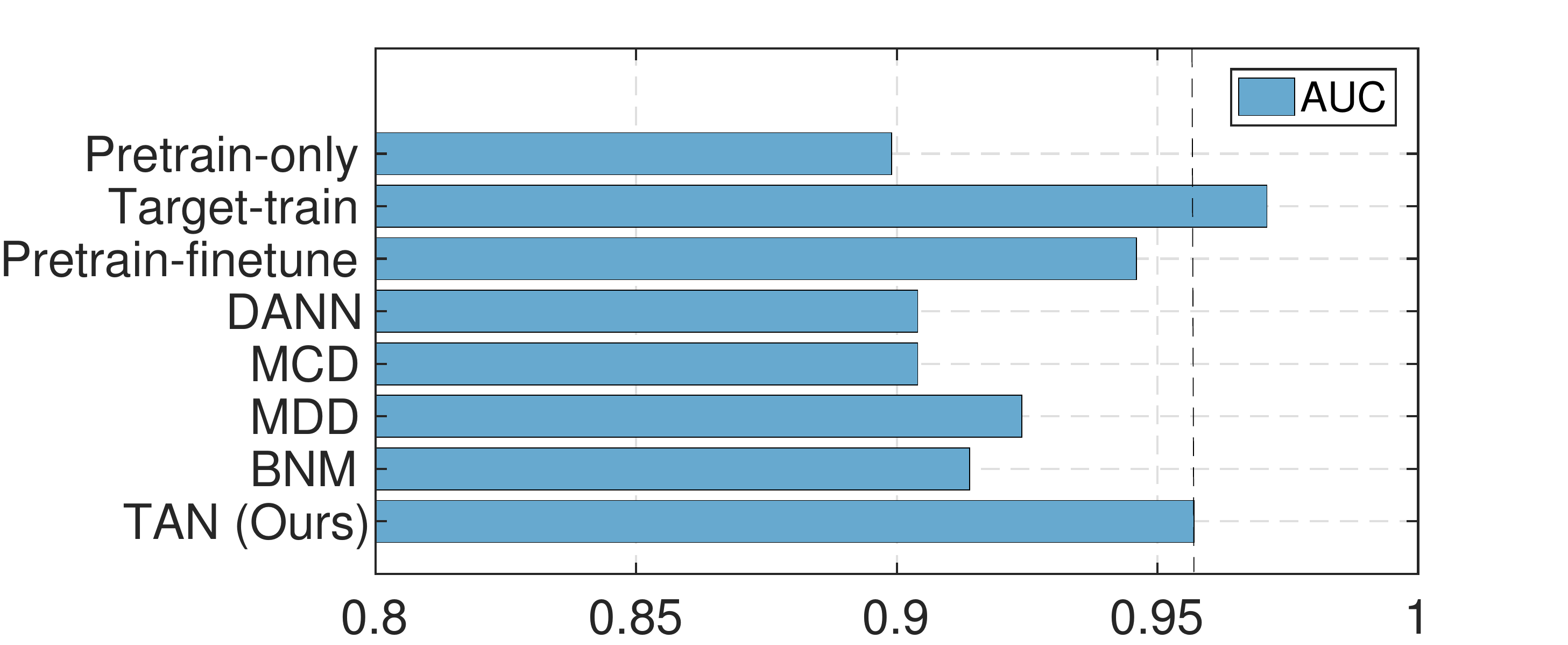}
	\caption{Area Under Curve (AUC) of different methods, indicating that the proposed \method achieves the best AUC scores in all comparison methods other than the ideal state (target-train)}
	\label{fig-auc}
\end{figure}

We also compute the AUC (Area Under Curve) of our method and compare it with other baselines in \figurename~\ref{fig-auc}.
The results show that other than target-train, which is the ideal state that uses the target domain labels for training, our method achieves the best AUC values compared to all other baselines.

\begin{table*}[t!]
	\centering
	\caption{Accuracy (\%) on Office-Home dataset (ResNet-50)}
	\label{tb-officehome}
	\resizebox{.8\textwidth}{!}{
		\begin{tabular}{cccccccccccccc}
			\toprule
			Method & A$\rightarrow$C & A$\rightarrow$P & A$\rightarrow$R & C$\rightarrow$A & C$\rightarrow$P & C$\rightarrow$R & P$\rightarrow$A & P$\rightarrow$C & P$\rightarrow$R & R$\rightarrow$A & R$\rightarrow$C & R$\rightarrow$P & AVG \\ \hline
			ResNet & 34.9 & 50.0 & 58.0 & 37.4 & 41.9 & 46.2 & 38.5 & 31.2 & 60.4 & 53.9 & 41.2 & 59.9 & 46.1 \\ 
			DAN~\cite{long2015learning} & 43.6 & 57.0 & 67.9 & 45.8 & 56.5 & 60.4 & 44.0 & 43.6 & 67.7 & 63.1 & 51.5 & 74.3 & 56.3 \\ 
			DANN~\cite{ganin2014unsupervised} & 45.6 & 59.3 & 70.1 & 47.0 & 58.5 & 60.9 & 46.1 & 43.7 & 68.5 & 63.2 & 51.8 & 76.8 & 57.6 \\ 
			JAN~\cite{long2017deep} & 45.9 & 61.2 & 68.9 & \textbf{50.4} & 59.7 & 61.0 & 45.8 & 43.4 & 70.3 & 63.9 & \textbf{52.4} & \textbf{76.8} & 58.3 \\ 
			\method (Ours) & \textbf{46.9} & \textbf{61.3} & \textbf{71.4} & 49.0 & \textbf{62.8} & \textbf{63.9} & \textbf{53.6} & \textbf{48.8} & \textbf{73.4} & \textbf{66.9} & 45.4 & 75.2 & \textbf{59.9} \\ \bottomrule
		\end{tabular}
	}
\end{table*}

\section{Design criteria of task semantic adaptor}
In this section, we pay special attention to the design criteria of the task semantic adaptation network $M_\theta$.
We extensive analyze the following aspects: 1)~network structure, 2)~training criterion, and 3)~activation function.
Thorough analysis of these properties is valuable for designing better network and learning strategies in similar problems to ensure better performance.

\subsection{Network structure}
We design different structures of $M_\theta$ and record its performance in \tablename~\ref{tb-structure}.
As shown in the results, different structures produce different results and all results are better than comparison methods in \tablename~2 of the main paper.
This means that the MLP structure of $M_\theta$ is effective.
In general, complex structure is better at learning meaningful representations.
In contrast, a simple structure may be worse in feature learning, but more difficult to overfit.
Based on our experiments, we choose the structure $\mathrm{in}-128-64-1$ for $M_\theta$.

\begin{table}[h]
	\centering
	\caption{Different structures of $M_\theta$}
	\label{tb-structure}
	\begin{tabular}{cccc}
		\toprule
		Structure & P (\%) & R (\%) & F1 (\%) \\ \hline
		$\mathrm{in}-512-256-1$ & 70.2 & 66.7 & 68.4 \\ 
		$\mathrm{in}-256-128-1$ & 69.0 & 66.7 & 67.8 \\ 
		$\mathrm{in}-128-64-1$ & 75.0 & 65.0 & 69.6 \\ 
	 \bottomrule
	\end{tabular}
\end{table}

\subsection{Training criterion}

\begin{figure}[htbp]
	\centering
	\includegraphics[width=.25\textwidth]{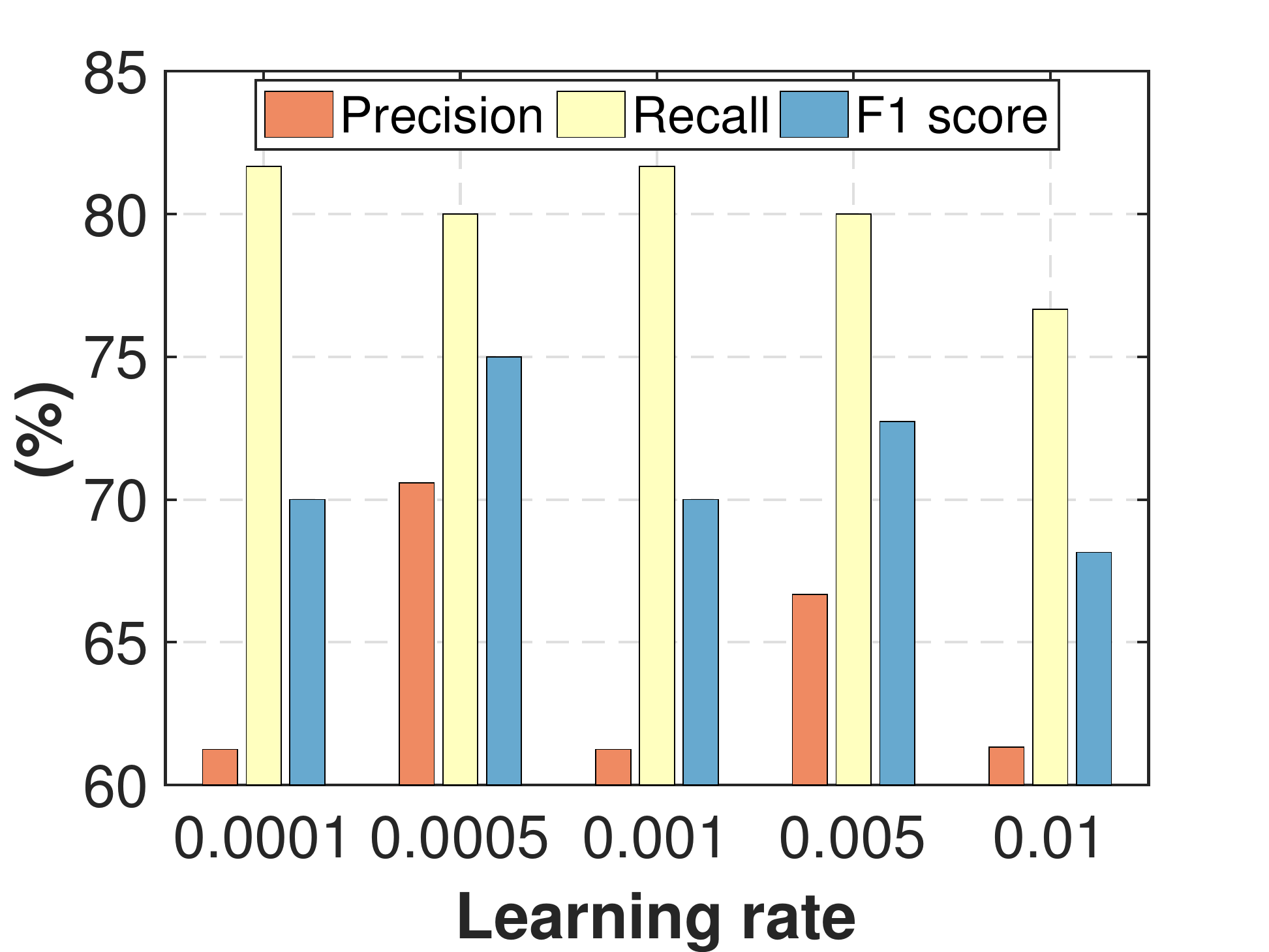}
	\caption{Different learning rates of $M_\theta$}
	\label{fig-lr}
	\vspace{-.1in}
\end{figure}

With regarding to the training criterion, we select different learning rates for the task semantic adaptor $M_\theta$ in $[0.0001, 0.0005, 0.001, 0.005, 0.01]$ and record the performance in \figurename~\ref{fig-lr}.
The results indicate that $M_\theta$ can achieve similar performance with different learning rates.
Therefore, to achieve the best performance we use the learning rate $0.0005$.

\subsection{Activation function}
We show the performance of different activation functions (i.e., $\sigma$ in Eq. (7) of the main paper) in \tablename~\ref{tb-activation}.
The results show that $\mathrm{tanh}$ can generally lead to better performance for our problem while $\mathrm{sigmoid}$ can also achieve good results.
Therefore, we use $\mathrm{tanh}$ in this work.

\begin{table}[h]
	\centering
	\caption{Different activation functions for $M_\theta$}
	\label{tb-activation}
	\begin{tabular}{cccc}
		\toprule
		Activation func & P (\%) & R (\%) & F1 (\%) \\ \hline
		$\mathrm{tanh}$            & 70.6 & 80.0    & 75.0    \\ 
		$\mathrm{sigmoid}$         & 62.8 & 81.7 & 71.0 \\ 
		$\mathrm{softplus}$        & 48.6  & 86.7 & 62.3 \\ 
		$\mathrm{relu}$            & 67.2 & 71.7 & 69.4 \\ \bottomrule
	\end{tabular}
	\vspace{-.1in}
\end{table}

%
%
%
%

\section{Detailed information of experiments}

For Bacterial and Viral datasets, the detailed information on source and target domain split in \tablename~\ref{tb-dataset-split}.
The split information of COVID-DA is omitted which can be seen in \cite{zhang2020covidda}
It is clear that the COVID-19 samples are not seen during training, which makes our unsupervised task transfer problem really challenging.

We also note that although the number of viral pneumonia samples is smaller than bacterial pneumonia samples, the transfer learning performance on viral dataset is almost the same as bacterial dataset (cf. \tablename~3 and 4 in the main paper, 95.2 vs. 95.4 F1 score).
This maybe due to the high similarity between viral pneumonia and COVID-19 pneumonia as COVID-19 is also caused by a certain type of virus called ``SARS-CoV-2''. Again, this reflects the fact that similarity matters in transfer learning.

\begin{table}[htbp]
	\caption{Detailed information on domain split of Bacterial and Viral datasets.}
	\label{tb-dataset-split}
	\resizebox{.48\textwidth}{!}{
	\begin{tabular}{ccccccc}
		\toprule
		Dataset & Domain & Sympotom & \#normal & \#pneumonia & \#COVID-19 & \#Total \\ \hline
		\multirow{3}{*}{Bacterial} & source & Pneumonia & 1,660 & 3,001 & 0 & 4,661 \\  
		& target & COVID-19 & 1,610 & 0 & 1,281 & 2,891 \\ \hline
		\multirow{3}{*}{Viral} & source & Pneumonia & 1,660 & 1,656 & 0 & 3,316 \\
		& target & COVID-19 & 1,610 & 0 & 1,281 & 2,891 \\ \bottomrule
	\end{tabular}
}
\vspace{-.1in}
\end{table}

\section{Results on standard DA datasets}

The results on Office-Home~\cite{venkateswara2017deep} dataset is shown in \tablename~\ref{tb-officehome}, and the results on VisDA-2017 dataset that uses all the classes are in \tablename~\ref{tb-visda-all}.
Note that \method does not focus on the standard domain adaptation tasks, thus, we do not compare it with the latest DA methods.
These results show that although the proposed \method is not tailored for traditional domain adaptation tasks, it still achieves competitive results compared to several strong baselines.
In the future, it is available to develop new \method-based methods for traditional DA tasks to increase its performance.

\begin{table}[htbp]
\centering
\caption{Accuracy (\%) on VisDA-2017 dataset (ResNet-50)}
\label{tb-visda-all}
\resizebox{.2\textwidth}{!}{
\begin{tabular}{cc}
\toprule
Method & syth $\rightarrow$ real \\ \hline
ResNet~\cite{he2016deep} & 52.4 \\ 
DAN~\cite{long2015learning} & 61.1 \\ 
DANN~\cite{ganin2014unsupervised} & 57.4 \\ 
JAN~\cite{long2017deep} & 61.6 \\ 
\method (Ours) & \textbf{64.0} \\ \bottomrule
\end{tabular}
}
\end{table}

\section{Parameter sensitivity analysis}

\begin{figure}[htbp]
	\centering
	\includegraphics[width=.25\textwidth]{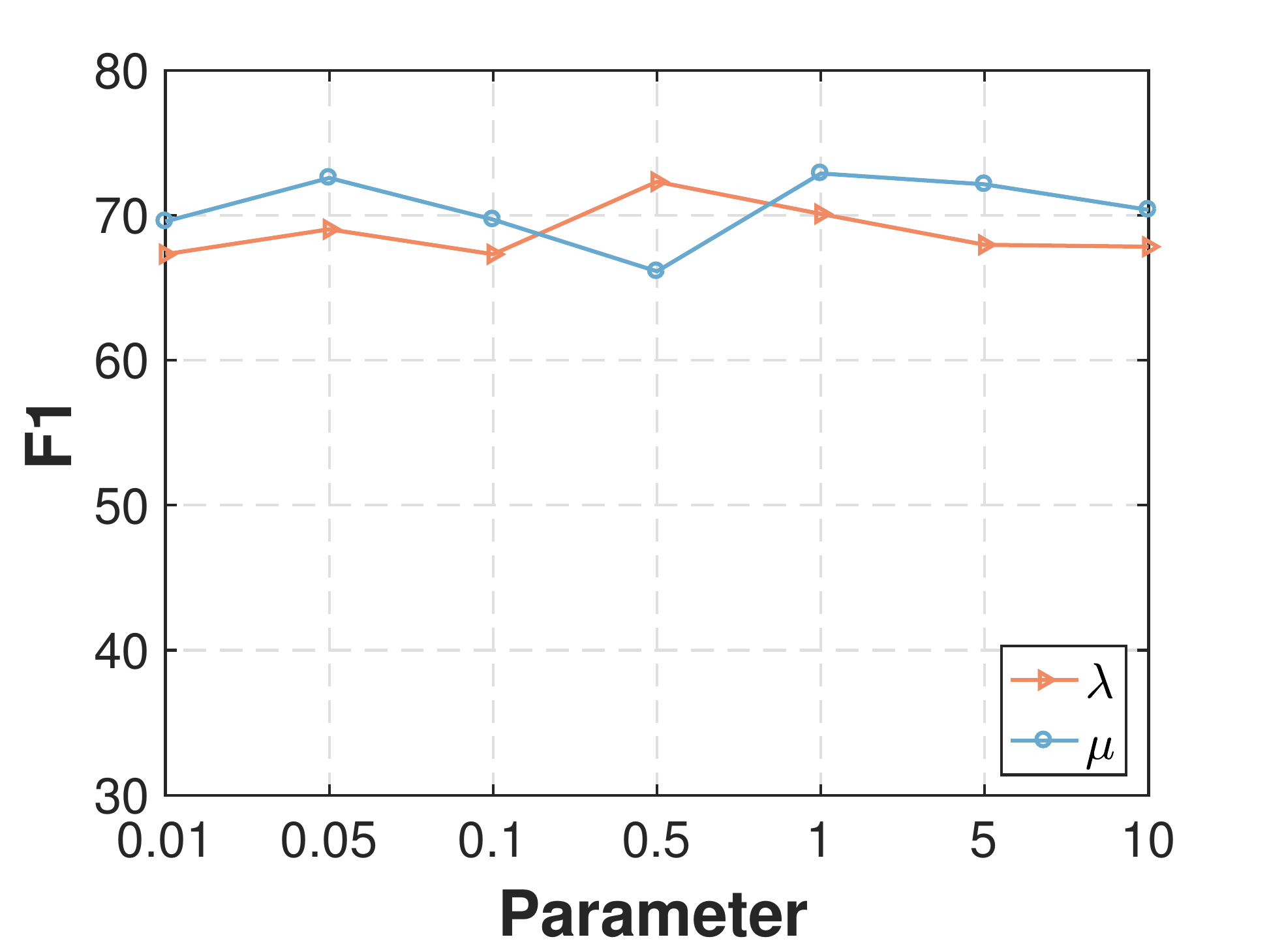}
	\caption{Parameter sensitivity of $\lambda$ and $\mu$}
	\label{fig-para}
\end{figure}

We show in \figurename~\ref{fig-para} that the two parameters $\lambda, \mu$ are relatively robust to different values, making our proposed method easy applicable to real applications.

